\documentclass[%
 reprint,
 amsmath,amssymb,
 aps,
]{revtex4-2}

\usepackage{graphicx}
\usepackage{dcolumn}
\usepackage{bm}

\usepackage{algorithmic}
\usepackage{algorithm}
\usepackage{color}
\usepackage{amsmath}
\usepackage{graphicx}
\usepackage{caption}
\usepackage{amsmath}
\usepackage{subfigure}

\usepackage[justification=centering]{caption}

\usepackage{marvosym}

\begin{document}

\preprint{APS/123-QED}

\title{Enhancing Graph Topology and Clustering Quality: A Modularity-Guided Approach}

\author{Yongyu Wang\textsuperscript{\Letter}}
 \email{wangyongyu1@jd.com}
\author{Shiqi Hao}%
 \author{Xiaoyang Wang}%
 \author{Xiaotian Zhuang}%
\affiliation{%
 JD Logistics\\
}%

\date{\today}

\begin{abstract}

Current modularity-based community detection algorithms attempt to find cluster memberships that maximize modularity within a fixed graph topology. Diverging from this conventional approach, our work introduces a novel strategy that employs modularity to guide the enhancement of both graph topology and clustering quality through a maximization process. Specifically, we present a modularity-guided approach for learning sparse graphs with high modularity by iteratively pruning edges between distant clusters, informed by algorithmically generated clustering results. To validate the theoretical underpinnings of modularity, we designed experiments that establish a quantitative relationship between modularity and clustering quality. Extensive experiments conducted on various real-world datasets demonstrate that our method significantly outperforms state-of-the-art graph construction methods in terms of clustering accuracy. Moreover, when compared to these leading methods, our approach achieves up to a hundredfold increase in graph construction efficiency on large-scale datasets, illustrating its potential for broad application in complex network analysis.

\end{abstract}

\maketitle


\section{Introduction}

Graph-based methodologies are pivotal in numerous machine learning and data mining tasks, owing to the graph's inherent strength in encapsulating the intricate structures and interrelationships within data sets. The efficacy of a graph-based approach hinges on the integrity of the graph itself, as the quality of the graph profoundly influences the algorithm's solution quality. Over the past decades, a variety of graph construction (or learning) techniques have been introduced.

The \(k\)-nearest neighbor (\(k\)-NN) graph, for instance, is widely adopted due to its simplicity and effectiveness in capturing the local manifold structure of data. In a \(k\)-NN graph, each node is linked to its \(k\) nearest neighbors, which helps to maintain robustness against outliers. Nevertheless, the fixed-size neighborhood used in \(k\)-NN graphs can restrict their ability to represent the global manifold structure adequately. The \(\epsilon\)-neighborhood graph offers an alternative by connecting each node to all other nodes within a specified distance \(\epsilon\). However, selecting an appropriate \(\epsilon\) value is challenging and sometimes impractical, especially when clusters within the data vary significantly in size. \cite{premachandran2013consensus} suggested enhancing the \(k\)-NN graph by extracting consensus information to prune noisy edges, where edges with a consensus value below a certain threshold are eliminated. While this method mitigates the effect of noisy connections, it may also inadvertently discard valuable structural information. More recent research has explored the use of graph signal processing (GSP) techniques to refine graph learning methods. These approaches aim to estimate sparse graph Laplacians more accurately. For example, \cite{egilmez2017graph} introduced a method that constrains the precision matrix to be a graph Laplacian and maximizes the posterior estimate of a Gaussian Markov Random Field (GMRF), incorporating an \(l1\)-regularization term to maintain graph sparsity. Additionally, \cite{kalofolias2017large} developed an approach that employs approximate nearest-neighbor methods to decrease the number of variables in the optimization process, thereby enhancing computational efficiency. However, these advanced Laplacian estimation techniques typically require computational time on the order of \(O(n^2)\) for each iteration, which poses a significant challenge for their application in large-scale real-world problems. Therefore, efficiently constructing a high-quality graph remains a challenging task.

In this paper, we propose leveraging modularity to enhance the quality of the graph. Modularity is a key concept in community detection algorithms. In these algorithms \cite{newman2006modularity,blondel2008fast,traag2019louvain}, for a given community network, the aim is to find a clustering division that maximizes modularity as the outcome of community partitioning. Unlike the conventional use of modularity in these methods, we introduce a framework that utilizes modularity to optimize the graph's topological structure. Specifically, we use modularity to iteratively identify and remove redundant and incorrect edges. In our framework, the quality of graph-based clustering algorithms and the graph's topological structure mutually reinforce each other, improving alternately. When the iterative process can no longer increase modularity, we output the final graph topology. Running graph clustering algorithms on this optimized graph topology can significantly enhance clustering accuracy. Experiments on multiple real-world benchmark datasets of large scale indicate that, compared to state-of-the-art graph construction methods, our approach can build graph topologies that significantly improve the accuracy of graph clustering algorithms. Moreover, our graph construction is hundreds of times more efficient than the state-of-the-art methods.

Another contribution of this article is the empirical quantification of the relationship between modularity and clustering accuracy through the method we propose, thus substantiating the modularity theory. In community detection tasks, most clustering divisions lack a ground truth, or they involve multi-label partitions with extremely complex overlap relationships\cite{yang2014structure,harenberg2014community}. In many research papers on community detection, the value of modularity is directly used as the quantitative evaluation \cite{newman2006modularity,blondel2008fast,traag2019louvain}, which is clearly not very persuasive. Therefore, how to quantitatively evaluate the effectiveness of community detection algorithms remains a highly controversial issue, with no consensus reached thus far \cite{yang2014structure,harenberg2014community}. This paper introduces a modularity-based graph clustering algorithm that enables the use of standard datasets with a unique ground truth label for each sample to obtain the quantitative relationship between modularity and clustering accuracy. Experimental results on the Pendigits dataset using our algorithm indicate that modularity and clustering accuracy increase synchronously.

\section{Preliminary}

\subsection{Modularity}\label{distortionPerspective}
Modularity, denoted by $Q$, is a metric used in community detection to assess the quality of a network division. It measures the strength of division of a network into communities by comparing the density of edges within communities to the density of edges between communities. The modularity $Q$ is defined as \cite{newman2006modularity}:

\begin{equation}\label{eqn:modularity}
Q = \frac{1}{2m} \sum_{ij} \left[ A_{ij} - \frac{k_i k_j}{2m} \right] \delta(c_i, c_j),
\end{equation}

where $A_{ij}$ represents the adjacency matrix of the network, with $A_{ij} = 1$ if there is an edge between nodes $i$ and $j$, and $A_{ij} = 0$ otherwise. The degree of node $i$ is represented by $k_i$, and $m$ is the total number of edges in the network. The Kronecker delta function $\delta(c_i, c_j)$ is $1$ if nodes $i$ and $j$ belong to the same community, and $0$ otherwise.

Community detection algorithms such as the Louvain Method \cite{blondel2008fast} and the Leiden Method \cite{traag2019louvain} leverage modularity to identify meaningful communities within complex networks. Modularity serves as a quality measure that quantifies the degree to which a network can be divided into distinct communities. These methods iteratively optimize modularity by reassigning nodes to communities, with the goal of maximizing the difference between observed and expected intra-community connections. This process results in the detection of cohesive groups of nodes that exhibit higher connectivity within their respective communities compared to what would be expected by chance. These algorithms efficiently uncover community structures, making them valuable tools in network analysis, social sciences, biology, and other domains where understanding network organization is essential.

\subsection{The Significance of Graph Topology in Clustering}

Graph topology plays a crucial role in detecting non-convex and linearly non-separable patterns. To illustrate this, we use the most fundamental clustering task in artificial intelligence as an example.

The widely adopted $k$-means algorithm, a non-graphical clustering method, primarily relies on distance metrics to minimize the total distances between data points and their respective cluster centroids. However, it often falls short in providing satisfactory clustering results for complex datasets, exemplified by well-known scenarios like the two moons and two circles datasets. 

In contrast, graph-based algorithms, such as spectral clustering, leverage the connectivity information embedded in the graph topology. This allows them to overcome the limitations of $k$-means and generate more accurate clustering results for complex, non-linear data distributions.

\section{Methods}
\subsection{Overview}
In broad terms, our approach draws inspiration from recent modularity-based community detection methods. However, it deploys modularity in a distinct manner to accomplish a unique objective. Unlike existing methods that employ modularity to guide node clustering within a fixed and predefined graph topology, our method introduces a novel perspective.

Existing modularity-based algorithms such as Leiden and Louvain algorithms iteratively adjust the cluster membership of individual nodes, seeking to reach the point of maximal modularity. Once this maximum modularity is achieved, the algorithm terminates, and the resulting cluster memberships are considered the final clustering results. In contrast, our approach deviates from the conventional paradigm by redefining the role of modularity within the clustering process. Instead of the traditional approach of altering node cluster memberships within a fixed graph topology to maximize modularity, we aim to optimize the topology based on a given clustering result by maximizing modularity. Once we obtain an improved graph topology through this optimization process, running clustering algorithms on this enhanced topology enables us to achieve more accurate clustering results.

\subsection{Algorithm Details}

Let us delve deeper into the definition of modularity as given in (\ref{eqn:modularity}) in Section \ref{distortionPerspective}.

When edges between two clusters are removed, it directly affects the modularity calculation by altering both the adjacency matrix $A_{ij}$ and the total number of edges $m$. Generally, removing inter-cluster edges tends to increase modularity since modularity rewards network structures with dense intra-cluster connections and sparse inter-cluster connections.

More specifically, removing edges between different communities reduces the $\frac{k_i k_j}{2m}$ term in the equation because $m$ (the total number of edges) decreases, and since $\delta(c_i, c_j) = 0$ for nodes in different communities, no subtraction of actual existing edge contributions occurs. Consequently, the removal of edges between communities usually leads to an increase in modularity, reflecting a more distinct community structure. However, there is a limit to this increase. If too many edges are removed, the network may become too fragmented, and communities may become disconnected, at which point modularity may no longer be a good measure. Therefore, removing edges to increase modularity should be done without compromising the connectivity within communities.

Based on the above analysis, we propose utilizing modularity to direct the optimization of graph topology and the enhancement of clustering accuracy.

The initial challenge in our method is that at the beginning, we neither have a sufficiently optimized topology nor highly accurate clustering results. If we proceed to remove edges between different clusters based on inaccurate cluster memberships, this will not only fail to optimize the topology but may also result in the removal of numerous correct edges, thereby worsening the topology. To address this, for each cluster, we calculate the centroid and identify the cluster that is furthest away by comparing the distances to the centroids of all other clusters, and then we remove all edges between it and the cluster that is furthest away. Although the initial accuracy of the clustering results may not be high, the points within a cluster and those in the cluster that is furthest away are still very likely to not belong to the same class. Therefore, removing the edges between them is a safe and confident approach.

Subsequently, after removing the edges, we rerun the clustering algorithm on this optimized topology. The clustering algorithm should yield more accurate results on the optimized topology. Then, based on this more accurate clustering result, we search for the furthest cluster from each cluster and remove the edges between them.

Based on the aforementioned process, the quality of the graph topology and the accuracy of graph-based clustering enhance each other, improving iteratively.

After several iterations of repeating the above process, our method will output the final graph once the modularity stabilizes, indicating that further changes to the graph's topology will not result in significant improvements in modularity.

\subsection{Complexity Analysis}

The time complexity is dominated by the clustering algorithm employed. It should be noted and emphasized that our method is not designed for any specific clustering algorithm, but rather it is a generalized optimization approach applicable to all graph-based clustering algorithms. Currently, the latest research progress have reduced the time complexity of representative graph-based clustering algorithms, such as spectral clustering, to near-linear \cite{wang2022improving}. Therefore, if we use the spectral clustering algorithm for clustering in Algorithm \ref{alg:flow}, the complexity of Algorithm \ref{alg:flow} is near-linear.

\section{Experiment}

In this paper, we construct the initial graph topology using the most commonly employed $k$-Nearest Neighbors ($k$-NN) method and perform clustering with the classic spectral clustering algorithm. The value of $k$ is set to 10 in the k-NN method. Experiments are performed using MATLAB running on the Laptop.

\subsection{Data Sets}
We have conducted experiments on the following three real-world benchmark data sets:\\

\textbf{COIL-20} includes 1,440 gray scale images of 20 different objects and each image is represented by $1,024$ attributes; \\

\textbf{PenDigits} includes  $7,494$ handwritten digits and each digit is represented by $16$ attributes;\\

\textbf{USPS} includes   $9,298$ images of USPS hand written digits with $256$ attributes.

\subsection{Comparison Methods}
We compare the proposed method against both the baseline and the state-of-the-art graph learning (construction) methods by applying the same spectral clustering algorithm to the graphs constructed by these various methods and evaluating the accuracy of the clustering results. The methods being compared include:\\

1) $k$-NN graph: the most widely used graph construction method. Each node is connected to its $k$ nearest neighbors.\\

2) Consensus $k$-NN graph: the state-of-the-art graph edge selection methods for improving the performance of $k$-NN graph. It improves the robustness of the $k$-NN graph by using the consensus information from different neighborhoods of a given $k$-NN graph.\\

3) LGSS: the most recent progress in graph learning field from the GSP perspective. It can automatically select the parameters of the model for achieving the desired graph properties. 

\subsection{Evaluation Metric}
We measure the quality of clustering with two metrics: clustering accuracy (ACC) and normalized mutual information (NMI) between the clustering results generated by clustering algorithms and the ground-truth labels provided by the data sets. The two metrics are defined as follows:

\subsubsection{Clustering Accuracy}
The clustering accuracy is defined as:
\begin{equation}\label{eqn:scale}
ACC= \frac{\sum\limits_{j = 1}^n  {\delta {(y_i,map(c_i))}}}{{n}},
\end{equation}
where $n$ is the number of data instances in the data set, $y_i$ is the ground-truth label provided by the data set,and $C_i$ is the label generated by the clustering algorithm. $\delta (x,y)$ is a delta function defined as: $\delta (x,y)$=1 for $x=y$, and $\delta (x,y)$=0,  otherwise.
 $map(\bullet)$ is a permutation function that maps each cluster index $c_i$  to a ground truth label, which can be realized using the Hungarian algorithm \cite{papadimitriou1998combinatorial}. A higher value  of $ACC$ indicates better  clustering quality.

\subsubsection{Normalized Mutual Information}
For two random variables $P$ and $Q$, normalized mutual information is defined as \cite{strehl2002cluster}:
\begin{equation}\label{eqn:scale}
NMI= \frac{I(P,Q)}{{\sqrt{H(P)H(Q)}}},
\end{equation}
where $I(P,Q)$ denotes the mutual information between $P$ and $Q$, while $H(P)$ and $H(Q)$ are entropies of $P$ and $Q$. In practice, the NMI metric can be calculated as follows \cite{strehl2002cluster}:
\begin{equation}\label{eqn:scale}
NMI= \frac{{\sum\limits_{i = 1}^k}{\sum\limits_{j = 1}^k}{n_{i,j}}\log(\frac{{n}\cdot{n_{i,j}}}{{n_i}\cdot{n_j}}) }{{\sqrt{(\sum\limits_{i = 1}^k {{n _{i}}{\log\frac{n_i}{n}}})(\sum\limits_{j = 1}^k {{n _{j}}{\log\frac{n_j}{n}}})}}},
\end{equation}
where $n$ is the number of  data points in the data set, k is the number of clusters, $n_i$ is the number of data points in cluster $C_i$ according to the clustering result generated by algorithm, $n_j$ is the number of data points in class $C_j$ according to the ground truth labels provided by the data set, and $n_{i,j}$ is the number of data points in cluster $C_i$ according to the clustering result as well as in class $C_j$ according to the ground truth labels. The NMI value is in the range of [0, 1], while a higher NMI value indicates a better matching between the algorithm generated result and ground truth result.\\

\vspace{-0.1in}

\subsection{Clustering Quality Results}

We perform spectral clustering algorithm on the graphs generated by the four graph construction methods. The clustering accuracy results and NMI results are shown in Table~\ref{table:compare3} and Table~\ref{table:compare4}, respectively.

As shown in Table~\ref{table:compare3}, the proposed method can consistently lead to dramatic performance improvement over the given graph. By applying our optimization method on $k$-NN graph, it achieves more than $6\%$, $12\%$ and $16\%$ clustering accuracy gains on the three data sets, respectively. For the Pendigits dataset, our method has $12\%$ clustering accuracy gain over the second-best method. The superior clustering results clearly demonstrate the effectiveness of the proposed method. As shown in Table~\ref{table:compare4}, for both the Pendigits and the USPS data sets, our method provides best NMI results among all the compared methods.

\begin{table*}[!htbp]\label{compare3}
\begin{center}

\caption{Clustering Accuracy (\%)}
\scalebox{1.4}{
\begin{tabular}{ |c|c|c|c|c|c|c|c|c|c|c|c }
\hline
 
 \hline   Data Set &$k$-NN &Consensus&LGSS&Our method\\
 \hline   COIL20  &75.72&81.60 &85.49&82.22\\
 \hline   PenDigits &74.36&71.08 &74.53&86.42 \\
 \hline   USPS     &64.31&68.54 &81.50&80.55 \\
 
 \hline
\end{tabular}}\label{table:compare3}
\end{center}
\end{table*}

\begin{table*}[!htbp]\label{compare4}
\begin{center}

\caption{NMI}
\scalebox{1.4}{
\begin{tabular}{ |c|c|c|c|c|c|c|c|c|c|c|c }
\hline
 
 \hline   Data Set &$k$-NN &Consensus&LGSS&Our method\\
 \hline   COIL20  &0.86&0.90&0.95&0.88\\
 \hline   PenDigits &0.79&0.79 &0.77&0.81 \\
 \hline   USPS     &0.79&0.81 &0.84&0.85 \\
 
 \hline
\end{tabular}}\label{table:compare4}
\end{center}
\end{table*}

\subsection{Graph Construction Time Results}

To assess the efficiency of graph construction, we report the time required for building graphs with both our method and current state-of-the-art methods, as shown in Table~\ref{table:compare5}.

\begin{table*}[!htbp]\label{compare5}
\begin{center}

\caption{Graph learning (construction) time (Seconds)}
\scalebox{1.57}{
\begin{tabular}{ |c|c|c|c|c|c|c|c|c|c|c|c }

 \hline   Data Set &Consensus&LGSS&Our method\\
 \hline   COIL20  &2.43 &13.56&6.32 \\
 \hline   PenDigits &172.51 &1085.43&7.87 \\
 \hline   USPS     &574.28 &2074.78&5.01 \\
 \hline
\end{tabular}}\label{table:compare5}
\end{center}
\end{table*}

It can be seen that our method is much more efficient compared to other methods. For the Pendigits data set, our method achieved \textbf{22X} and \textbf{138X} times speedup over the Consensus method and the LGSS method, respectively. For the USPS data set, our method achieved \textbf{115X} and \textbf{415X} times speedup over the Consensus method and the LGSS method, respectively. These results indicate that, compared with state-of-the-art graph construction methods, our graph optimization approach not only yields higher-quality graphs that significantly enhance clustering accuracy but also greatly improves efficiency.

\subsection{Quantitative Empirical Validation of the Relationship Between Modularity and Clustering Quality}

Although many algorithms in the field of community detection optimize clustering based on modularity, to the best of our knowledge, there have been no empirical quantitative results regarding the relationship between modularity and the quality of clustering to date. In this paper, we design experiments based on our proposed method and provide quantitative evidence confirming the relationship between modularity and clustering effectiveness.

\begin{figure*}[!h]
\centering\includegraphics[scale=0.35]{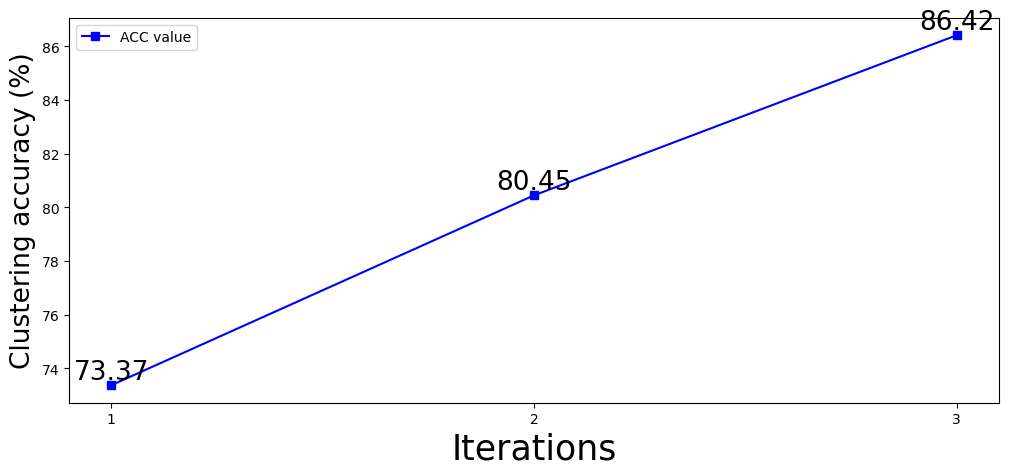}
\caption{Clustering accuracy through iterations of the Pendigits data set.\protect\label{fig:accvsiter.png}}
\end{figure*}

\begin{figure*}[!h]
\centering\includegraphics[scale=0.35]{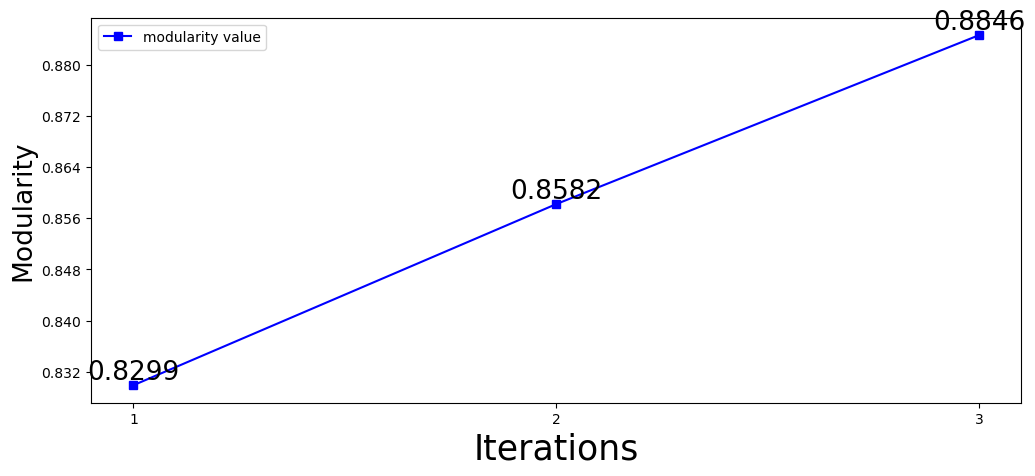}
\caption{Modularity through iterations of the Pendigits data set.\protect\label{fig:modularityvsiter.png}}
\end{figure*}

The clustering accuracy and modularity results through iterations of our method have been shown in Figure \ref{fig:accvsiter.png} and Figure \ref{fig:modularityvsiter.png}, respectively. The clustering accuracy and modularity results over the iterations of our method are presented in Figure \ref{fig:accvsiter.png} and Figure \ref{fig:modularityvsiter.png}, respectively. It can be observed that the clustering accuracy and modularity increase concurrently, which substantiates the validity of the modularity theory.

\section{Conclusion}\label{sect:conclusions}
In this work, we present a modularity-guided graph topology optimization method. We show that the graph topology learning problem can be solved by iteratively identifying and removing redundant and misleading edges to increase the modularity of the graph. Unlike traditional modularity-based  approaches which focus on updating cluster-membership of samples to maximize the modularity, our method aims to optimize the graph topology through the modularity maximization process. When comparing with state-of-the-art graph
construction (learning) approaches, our approach is more efficient and leads to substantially improved solution quality of graph-based clustering. The paper also validates the modularity theory through quantitative empirical evidence of the relationship between modularity and clustering quality.

\section{Acknowledgments}

During the interview process for the algorithm engineer position at JD Logistics, Zhangxun Liu, acting as an interviewer, mentioned the Girvan-Newman algorithm. This reference piqued the interest of Dr.Yongyu Wang, who consequently began to explore and understand modularity-based algorithms. On March 10, 2023, Zhangxun Liu organized a seminar on community detection algorithms, where Shiqi Hao introduced the details of the Louvain and Leiden algorithms. This seminar proved to be a catalyst for Dr. Yongyu Wang, inspiring the conception of the idea for this paper. As the first author and corresponding author of the paper, Dr.Yongyu Wang extends sincerely gratitude to Zhangxun Liu for the aforementioned contributions. 

After careful consideration, it is more appropriate to express thanks to Zhangxun Liu in the acknowledgements section rather than to include him as a co-author.

\nocite{*}

\bibliography{apssamp}

\end{document}